\documentclass[twocolumn]{aastex62}
\usepackage{natbib}
\usepackage{float}
\usepackage{amsmath}
\graphicspath{{./}{figures/}}

\accepted{February 2024}
\submitjournal{ApJ}
\shorttitle{Updated Cas~A NS Velocity}
\shortauthors{Holland-Ashford et. al.}

\begin{document}

\title{Updated Proper Motion of the Neutron Star in the Supernova Remnant Cassiopeia~A}
\correspondingauthor{Tyler Holland-Ashford}
\email{tyler.holland-ashford@cfa.harvard.edu}
\author{Tyler Holland-Ashford}
\affil{Center for Astrophysics $|$ Harvard \& Smithsonian, 60 Garden St, Cambridge MA 02138, USA}
\author{Patrick Slane}
\affil{Center for Astrophysics $|$ Harvard \& Smithsonian, 60 Garden St, Cambridge MA 02138, USA}
\author{Xi Long}
\affil{Center for Astrophysics $|$ Harvard \& Smithsonian, 60 Garden St, Cambridge MA 02138, USA}

%\date{February 2024}

\begin{abstract}
In this paper, we present updated estimates of the velocity of the neutron star (NS) in the supernova remnant (SNR) Cassiopeia~A using over two decades of {\it Chandra} observations. We use two methods: 1.) recording NS positions from dozens of {\it Chandra} observations, including the astrometric uncertainty estimates on the data points but not correcting the astrometry of the observations, and 2.) correcting the astrometry of the 13 {\it Chandra} observations that have a sufficient number of point sources with identified Gaia counterparts. For method \#1, we observe a heliocentric velocity of 275 $\pm$ 121 km/s, with an angle of 177 $\pm$ 22 degrees east of north. For method \#2, we observe a heliocentric velocity of 436 $\pm$ 89 km s$^{-1}$ at an angle of 158 $\pm$ 12 degrees. Correcting for galactic rotation and the sun's peculiar motion decreases these estimates to 256 km s$^{-1}$ at 167$^\circ$ and 433 km s$^{-1}$ at 151$^\circ$, respectively. % these velocity estimates by 5--20 km s$^{-1}$ and the estimated kick direction by 10--15$^\circ$.} 
Both of our estimates match with the explosion-center-estimated velocity of $\sim$350 km s$^{-1}$ and the previous 10--year baseline proper motion measurement of 570 $\pm$ 260 km s$^{-1}$, but our use of additional data over a longer baseline has led to a smaller uncertainty by a factor of 2--3. Our estimates rule out velocities $\gtrsim$600 km s$^{-1}$ and better match with simulations of Cassiopeia~A that include NS kick mechanisms.

\end{abstract}


\section{Introduction}
Neutron stars, the compact objects formed in some core-collapse supernovae (CCSNe), have typical observed velocities of hundreds of km s$^{-1}$ \citep{hobbs05,verbunt17,igoshev20}, with some sources reportedly reaching velocities of $\gtrsim$1000 km s$^{-1}$ (e.g., \citealt{chatterjee05}). These velocities are higher than can be explained from the disruption of a pre-explosion binary (v$_{\rm NS}\approx100$ km s$^{-1}$; \citealt{lai01}), indicating that supernova explosion processes are responsible for these kicks. These velocities can be compared to various supernova remnant (SNR) properties---in particular, ejecta asymmetries---to provide constraints on supernova progenitors and explosion processes.

The two prevailing theories for the origins of NS kicks invoke conservation-of-momentum like arguments with asymmetric supernova explosion forces. In the first theory, hydrodynamical instabilities during a supernova lead to the NS being kicked in the opposite direction as the bulk of ejecta \citep{scheck06, wongwathanarat13,janka17}. In the second theory, both ejecta and NS move in a direction opposite the bulk of anisotropic neutrino emission \citep{fryer06}.

Recent observations and theoretical predictions have provided stronger support for the first scenario. 3D simulations have generated kick velocities of up to $\sim$1000 km s$^{-1}$ in a direction opposite the bulk of ejecta (the ``Gravitational Tugboat Mechanism', see e.g., \citealt{wongwathanarat13,janka17,gessner18, burrows23}). Observational studies have  also confirmed that NSs are indeed preferentially kicked in a direction opposite the bulk ejecta motion in a sample of supernova remnants (SNRs) (\citealt{me17,katsuda18,bhalerao19}). As for the anisotropic neutrino emission scenario, although it is partially supported by an observed spin-kick alignment in some NSs \citep{johnston05}, generating kick velocities of $\sim$300 km s$^{-1}$ requires very strong ($\gtrsim$10$^{16}$ G) magnetic fields \citep{scheck06,wongwathanarat13}---unlikely, but still possible. Recent simulations showed that neutrino kicks of up to $\sim$100 km s$^{-1}$ are feasible and perhaps dominate over ejecta kicks in CCSNe with the lowest progenitor masses \citep{coleman22,burrows23}.

In order to use NSs as probes of SN explosion properties, it is vital to study NSs still embedded in SNRs. Such NSs can be directly linked to specific supernova explosions and compared to the SNR properties such as ejecta asymmetries, energy, distance, and age. However, these NSs are by definition young ($\lesssim$20~kyr old) and don't exhibit radio, infrared, or optical emission, and thus astronomers are forced to largely rely on X-ray telescopes to study them. A typical NS velocity of $\sim$400 km s$^{-1}$ at a distance of 3~kpc corresponds to sky motion of only $\sim$0.03" yr$^{-1}$. Only {\it Chandra} has sufficient spatial resolution to measure the proper motions of NSs, and doing so requires baselines of at least a decade for all but the closest objects. There are currently $\sim$20 velocity measurements of NSs embedded within SNRs, found using a mix of direct proper motion measurements (e.g., \citealt{auchettl15,temim17, shternin19, mayer21,long22}) and measuring the distance between NS positions and estimated SNR explosion sites (e.g., \citealt{winkler09, banovetz21}).

The SNR Cassiopeia~A is the youngest known CCSNR in the Milky Way, with an age of $\sim$350 years and a distance of 3.33 $\pm$ 0.1 kpc \citep{fesen06,alarie14}. Its proximity and strong ejecta emission has enabled detailed investigation of its ejecta and shock emission on small, sub-parsec scales. Optical observations have shown that Cas~A is an O-rich SNR (e.g., \citealt{chevalier78}), and light echoes from the explosion reveal that Cas~A was produced by an asymmetric Type IIb SN with variations in ejecta velocities of $\approx$4000 km s$^{-1}$ \citep{rest11}. The NS CXOU J232327.9$+$584842 was detected in the first-light image of Cas~A \citep{tananbaum99} and is a Central Compact Object (CCO), an odd class of neutron stars that are characterized by thermal emission, nonvariability, and weak magnetic fields (see e.g., \citealt{becker09,halpern10}).

There are already a few estimates of the velocity of the NS in Cas~A. \cite{thorstensen01} back-evolved the motion of various ejecta filaments to a central location, which should correspond to the origin of explosion (i.e., neutron star birth site). When combined with the SNR's age and current NS location, this gives an indirect estimate of the NS's velocity: $\sim$340 $\pm$ 30 km s$^{-1}$ at an angle of 168$^\circ$ $\pm$ 8$^{\circ}$ east of north (counterclockwise).

As for direct proper motion measurements, both \cite{delaney13} and \cite{mayer21} used {\it Chandra} {\sc HRC} data from 1999 to 2009 to measure the proper motion of the NS. The former study made use of one background source and 13 quasi-stationary flocculi to correct for astrometric offsets between observations, reporting a velocity of 390 $\pm$ 400 km s$^{-1}$. The latter authors combined same-epoch {\sc HRC} observations for more signal, detecting and using two background sources to find a velocity of 570 $\pm$ 260 km/s in a direction consistent with the explosion site measurement. The motivation of this paper is to use the additional 14 years of {\it Chandra} observation since 2009 to obtain a more precise NS velocity measurement.

In this paper, we present updated measurements of the proper motion of the CCO in Cas~A using 24 years of {\it Chandra} X-ray observations, calculated using multiple methods. First, we measure its velocity without using background point sources to correct for {\it Chandra} astrometry, simply making use of dozens of observations of Cas~A from 1999 to 2023. We include both detection uncertainty and astrometric uncertainty on the position of the NS at each epoch, fitting a line of best-fit to measure the overall motion. Using this method, each individual data point has large ($\gtrsim$0.5'') uncertainties, but the each additional data point enables a more precise measurement. In our second method, we use detected background point sources in observations of Cas~A to improve for {\it Chandra}'s astrometric accuracy. The precision of the NS's position at each epoch can be resolved down to $\sim$0.1'', but there are only a few observations with a sufficient number of detected calibration point sources with which to perform astrometry.

Our paper is formatted as follows: in Section~\ref{sec:methods}, we describe our data and the procedures used to reprocess and analyze it. In Section~\ref{sec:results}, we present both of our velocity estimates along with any caveats. In Section~\ref{sec:disc}, we discuss and compare our estimates to past works, both relating to Cas~A specifically and to identified SNR \& NS populations. 

\begin{deluxetable*}{lcccc||lcccc}
\tablecolumns{4}
\tablewidth{0pt} 
\tablecaption{{\it Chandra}  Observations\label{table:obslog}} 
\tablehead{ \colhead{ObsID} & \colhead{Date} &\colhead{Roll} &\colhead{Exposure}  & \colhead{Det} & \colhead{ObsID} & \colhead{Date} &\colhead{Roll} &\colhead{Exposure}  & \colhead{Det}\\
\colhead{} &\colhead{} &\colhead{Angle ($\deg$)} & \colhead{(ks)} & \colhead{} & \colhead{} &\colhead{} &\colhead{Angle ($\deg$)} & \colhead{(ks)} & \colhead{} }
\startdata
\multicolumn{5}{c}{{\sc ACIS} Observations (36 total; 32 usable)} & \multicolumn{5}{c}{{\sc HRC} Observations (29) }\\ \hline
114 & 2000-01-30 & 323.38 & 50 & ACIS-S & 171 & 1999-09-03 & 170.52 & 9 & HRC-I \\
1952 & 2002-02-06 & 323.38 & 50 & ACIS-S & 172 & 1999-09-05 & 172.69 & 9 & HRC-S \\
4634 & 2004-04-28 & 59.22 & 148 & ACIS-S & 1038\tablenotemark{c} & 2001-09-19 & 186.64 & 50 & HRC-S \\
4635 & 2004-05-01 & 59.22 & 138 & ACIS-S & 1505 & 1999-12-19 & 287.13 & 49 & HRC-I \\
4636 & 2004-04-20 & 49.77 & 150 & ACIS-S & 1549 & 2001-04-04 & 311.1 & 5 & HRC-I \\
4637 & 2004-04-22 & 49.77 & 170 & ACIS-S & 1550 & 2001-07-13 & 127.5 & 5 & HRC-I \\
4638 & 2004-04-14 & 40.33 & 170 & ACIS-S & 1857 & 2000-10-04 & 204.45 & 48 & HRC-S \\
4639 & 2004-04-25 & 49.77 & 80 & ACIS-S & 2781 & 2002-02-06 & 329.3 & 5 & HRC-I \\
5196 & 2004-02-08 & 325.5 & 50 & ACIS-S & 3705 & 2003-10-19 & 220.7 & 5 & HRC-I \\
5319 & 2004-04-18 & 49.77 & 40 & ACIS-S & 5157 & 2004-10-29 & 233.5 & 5 & HRC-I \\
5320 & 2004-05-05 & 65.14 & 54 & ACIS-S & 5164 & 2004-03-15 & 5.8 & 4 & HRC-I \\
6690\tablenotemark{a,b} & 2006-10-19 & 221.14 & 62 & ACIS-S & 6038 & 2005-10-23 & 225.9 & 5 & HRC-I \\
9117 & 2007-12-05 & 278.13 & 25 & ACIS-S & 6069 & 2005-04-12 & 36.9 & 5 & HRC-I \\
9773 & 2007-12-08 & 278.13 & 25 & ACIS-S & 6739 & 2006-03-22 & 12.8 & 5 & HRC-I \\
10935 & 2009-11-02 & 239.68 & 23 & ACIS-S & 6746 & 2006-10-15 & 215.9 & 5 & HRC-I  \\
10936 & 2010-10-31 & 236.48 & 31 & ACIS-S & 8370 & 2007-03-10 & 359.0 & 5 & HRC-I \\
12020 & 2009-11-03 & 239.68 & 22 & ACIS-S & 9700 & 2008-03-24 & 15.2 & 5 & HRC-I \\
13177 & 2010-11-02 & 236.48 & 17 & ACIS-S & 10227 & 2009-03-20 & 11.00 & 131 & HRC-S\\
13783\tablenotemark{a,b} & 2012-05-05 & 65.04 & 63 & ACIS-S & 10228 & 2009-03-28 & 24.86 & 130 & HRC-S\\
14229 & 2012-05-15 & 75.44 & 50 & ACIS-S  & 10229 & 2009-03-24 & 13.66 & 48  & HRC-S\\
14480 & 2013-05-20 & 75.14 & 50 & ACIS-S & 10698 & 2009-03-31 & 23.20 & 51 & HRC-S\\
14481 & 2014-05-12 & 75.14 & 50 & ACIS-S  & 10892 & 2009-03-26 & 13.66 & 124 & HRC-S \\
14482 & 2015-04-30 & 67.13 & 50 & ACIS-S & 11240 & 2009-12-20 & 287.02 & 12 & HRC-I\\
16946\tablenotemark{a,b} & 2015-04-27 & 54.97  & 68 & ACIS-S & 11955 & 2010-04-12 & 39.89 & 9 & HRC-I\\
17639\tablenotemark{a}  & 2015-05-01 & 67.13 & 43 & ACIS-S & 12057 & 2009-12-13 & 287.02 & 10 & HRC-I\\
18344 & 2016-10-21 & 214.20 & 26 & ACIS-S & 12058 & 2009-12-16 & 287.01 & 9 & HRC-I \\
19604 & 2017-05-16 & 76.58 & 50 & ACIS-S & 12059 & 2009-12-15 & 287.02 & 12 & HRC-I\\
19605 & 2018-05-15 & 75.23 & 49 & ACIS-S & 24840 & 2020-10-18 & 220.84 & 14 & HRC-I\\
19606 & 2019-05-13 & 75.14 & 49 & ACIS-S & 26440 & 2022-07-05 & 121.49 & 4 & HRC-I\\
19903 & 2016-10-20 & 214.20 & 25 & ACISS \\
22426\tablenotemark{a} & 2020-05-11 & 72.65  & 48 & ACIS-S \\
23248\tablenotemark{a} & 2020-05-14 & 72.65 & 28 & ACIS-S \\
26248\tablenotemark{b} & 2022-01-25 & 309.13  & 25 & ACIS-I \\
26655\tablenotemark{a} & 2022-12-30 & 297.17  & 29 & ACIS-S \\
27099\tablenotemark{a} & 2023-01-02 & 300.00  & 37 & ACIS-S \\
27100\tablenotemark{a} & 2023-01-15 & 310.11 & 27 & ACIS-S \\
\enddata
\tablenotetext{a}{Observations that don't use the entire chips. These observations are useful for measuring the NS position, but their reduced FoVs means they don't cover background sources.}
\tablenotetext{b}{These observations used high SIMZ offsets, which introduces significant astrometric errors and offsets, rendering such observations unsuitable for use in obtaining accurate point source positions.}
%\tablenotetext{c}{These observations used Exposure Mode: HRC Timing}
\end{deluxetable*}

\section{Observations and Data Analysis}
\label{sec:methods}
\subsection{{\it Chandra} Observations}
Cas~A was the target of {\it Chandra}'s first light image in 1999 \citep{hughes00}, and since then, {\it Chandra} has observed the SNR hundreds of times for over 3~Ms. However, most are calibration observations with observation times of a few kiloseconds: too short to measure a robust NS position, let alone fainter background sources for astrometric correction. We collected all {\sc ACIS} and {\sc HRC} imaging observations with observation times longer than a few kiloseconds, using the {\sc CIAO} version 4.13 tool \texttt{chandra\_repro} to reprocess the observations. Our list of observations---36 {\sc ACIS} and 29 {\sc HRC}---is reported in Table~\ref{table:obslog}. 

We removed observations with high chip offsets from our sample. Such offsets from the detector focal plane introduce additional astrometric uncertainties/offsets that would interfere with our ability to measure NS positions at each epoch. In total, we removed 4 observations with high ($\sim$10.7~mm) SIM-Z offsets---{\sc ACIS} ObsIDs 6690, 13783, 16946, and 26248.

\subsection{Point Source Detection}
We ran the {\sc CIAO} tool \texttt{wavdetect} on each of the observations in order to identify potential calibration point sources and then cross-matched detected point sources with known sources from the Gaia DR3 catalog\footnote{\url{https://www.cosmos.esa.int/web/gaia/dr3}} \citep{gaia16b,gaia23j,babusiaux23}. If there was spatial coincidence, we verified that: 1.) they remained coincident with Gaia counterparts across multiple epochs of observations, 2.) their spectra were distinct from ejecta emission, and 3.) they could be easily disentangled from surrounding SNR emission.

\begin{figure}
\begin{center}
\includegraphics[width=\columnwidth]{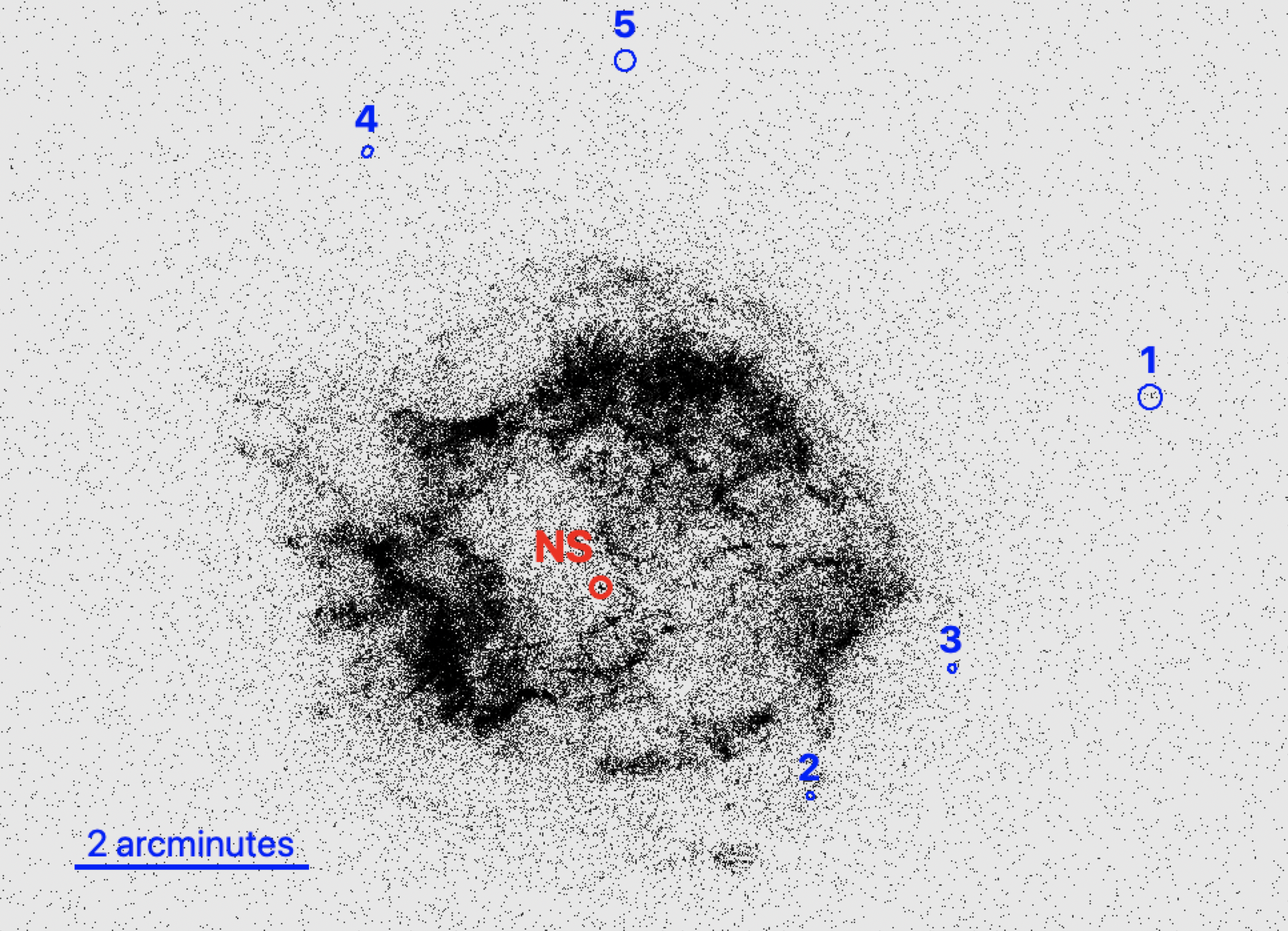}
\end{center}
\vspace{-5mm}
\caption{\footnotesize{A {\it Chandra} image of Cas~A, labeled with the NS (red) and detected point sources associated with Gaia counterparts (blue). The size of the circle around each point source is proportional to the uncertainty of the centroid position, i.e., how bright and well-resolved the point source is. The circle around the NS is greatly exaggerated for visibility.}}
\label{fig:pointsources}
\end{figure}

Across all observations, we detected 5 different point sources to use for astrometric correction. Figure~\ref{fig:pointsources} shows an image of Cas~A with the point sources' locations indicated, and Table~\ref{table:pointsources} lists their properties and the observations they are detected in. Although we detected 5 reference point sources to use for calibration, no single point source was present in all observations. We ended up using 13 observations---5 {\sc HRC} and 8 {\sc ACIS}--- that had a sufficient number of calibration point sources: $\ge$2 in most cases, but we loosen our restrictions to 1 for the latest observations taken in the past 5 years. 

\begin{deluxetable*}{rrrrrr}[!t]
\tablecolumns{7}
\tablewidth{0pt} 
\tablecaption{Detected Point Sources \label{table:pointsources}}
\tablehead{ \colhead{} & \colhead{1}  &\colhead{2} & \colhead{3} &\colhead{4} & \colhead{5}  }
\startdata
{\it Gaia} DR3 Source ID 20104... &  78284367990016 & 77356655102592 & 77253575869184 & 83743265755136 &  78868479720064  \\ 
{\it Gaia} RA (h:m:s; Epoch 2016) & 23:22:51.6674 & 23:23:14.1859 & 23:23:04.7861 & 23:23:43.2827 & 23:23:26.3270\\
{\it Gaia} Dec (d:m:s; Epoch 2016) & $+$58:50:19.7918 & $+$58:46:55.4287 & 58:48:00.0211 & 58:52:23.7984 & 58:53:10.3669\\
{\it Gaia} $\mu_\alpha$ (mas/yr)  & 15.64 $\pm$ 0.09 & 13.77 $\pm$ 0.02 & -1.29 $\pm$ 0.02 & 23.22 $\pm$ 0.01 & 2.08 $\pm$ 0.15 \\
{\it Gaia} $\mu_\gamma$ (mas/yr) & 0.3 $\pm$ 0.08 & -0.19 $\pm$ 0.02 & -1.64 $\pm$ 0.02 &  0.03 $\pm$ 0.01 & -1.02 $\pm$ 0.16   \\ 
Detected in {\sc HRC} ObsID \#1505?  & y & y & - & y & -\\
Detected in {\sc HRC} ObsID \#10892?  & y& y& -& y& -\\
Detected in {\sc HRC} ObsID \#10227?  & y& y& y& y& -\\
Detected in {\sc HRC} ObsID \#10228?  & y& y& -& -& -\\
Detected in {\sc HRC} ObsID \#24840? & y& -& -& -& -\\
Detected in {\sc ACIS} ObsID \#4634? & -& -& -& y& y\\
Detected in {\sc ACIS} ObsID \#4638? & -& -& -& y& y\\
Detected in {\sc ACIS} ObsID \#9773? & -& y& -& y& y\\
Detected in {\sc ACIS} ObsID \#10936? & -& y& y& -& -\\
Detected in {\sc ACIS} ObsID \#14482? & -& -& -& y& y\\
Detected in {\sc ACIS} ObsID \#19604? & -& -& -& y& -\\
Detected in {\sc ACIS} ObsID \#19605? & -& -& -& y& y\\
Detected in {\sc ACIS} ObsID \#19606? & -& -& -& y& -\\\hline
\enddata
\end{deluxetable*}

\subsection{{\it Chandra}'s Astrometric Uncertainties}
\label{subsec:astrouncert} 
The astrometric solution for any given {\it Chandra} image is only accurate to about $\sim$0.5--1.0'', depending on the detector and epoch of observation\footnote{\url{https://cxc.harvard.edu/cal/ASPECT/celmon/}}. To convert these reported values to 1-$\sigma$ individual uncertainties for RA \& Dec, we use the Rayleigh distribution formula:
\begin{equation}
    CDF = 1- e^{-x^2/(2\sigma^2)}
\end{equation}
The CDF---the Cumulative Distribution Function---is the reported confidence interval (often 68\% or 90\%), x is the reported angular circle uncertainty, and $\sigma$ is the 1D 1-$\sigma$ angular uncertainty value we are solving for.

\begin{deluxetable}{ccc}[!t]
\tablecolumns{7}
\tablewidth{0pt} 
\tablecaption{{\it Chandra} Astrometric Uncertainty \label{table:astro_uncert}} 
\tablehead{ \colhead{Detector}  &\colhead{2D 68\% CI} & 1D 1-$\sigma$ \\
\colhead{} &\colhead{(arcsec)} & \colhead{(arcsec)} }
\startdata
\multicolumn{3}{c}{2018-2023} \\ \hline
ACIS-S &  0.78'' & 0.52''  \\ 
ACIS-I & 0.65'' & 0.40''\\
HRC-S  & 0.93'' & 0.62''\\
HRC-I  & 0.71'' & 0.47'' \\ \hline
\enddata
%\tablenotetext{a}{Test}
%\vspace{-12mm}
\end{deluxetable}
%\vspace{-8mm}

We followed the advice of Lead Flight Director Dr. Tom Aldcroft to model the astrometric uncertainty evolution as a constant of $\sim$0.6'' (90\% confidence interval) prior to 2010 and linearly increasing to the reported 2018--2023-averaged values (see Table~\ref{table:astro_uncert}). We extrapolate these detector-specific uncertainties from 1999--2023, setting the individual detector values in 2010 to a factor of 0.6/1.11 smaller since the average 90\% confidence interval for 2021 is $\sim$1.11''.

\subsection{PSF Modeling and Fitting}
\label{subsec:psf}
To obtain astrometric corrections for each observation, we first needed to obtain precise point source positions that took the detector PSF into account. For this, we followed the methods of past papers (e.g., \citealt{mayer20,long22}). We produce PSF simulations using the {\sc CIAO} Chandra Ray Tracer (ChaRT) tool\footnote{\url{https://cxc.harvard.edu/ciao/PSFs/chart2/index.html}}, generating five iterations of each point source with a monochromatic spectrum of 1 keV with 0.01 photons cm$^{-2}$ s$^{-1}$. Most of our point sources are too faint for a robust spectrum to be extracted and modeled, and previous papers (e.g., \citealt{long22}) determined that using a model fit to the extracted spectrum didn't significantly improve the fits over using this simple spectrum. We then reprocessed those ChaRT output rayfiles using the \texttt{CIAO} tool \texttt{simulate\_psf}, which ran them through \texttt{MARX}\footnote{\url{https://space.mit.edu/CXC/MARX/}} and generated event files. We binned the resulting event files---{\sc HRC} observations by 1/2 and {\sc ACIS} observations by 1/8---to produce final pixel sizes of $\sim$0.065. 

\begin{figure*}
\begin{center}
\includegraphics[width=0.48\textwidth]{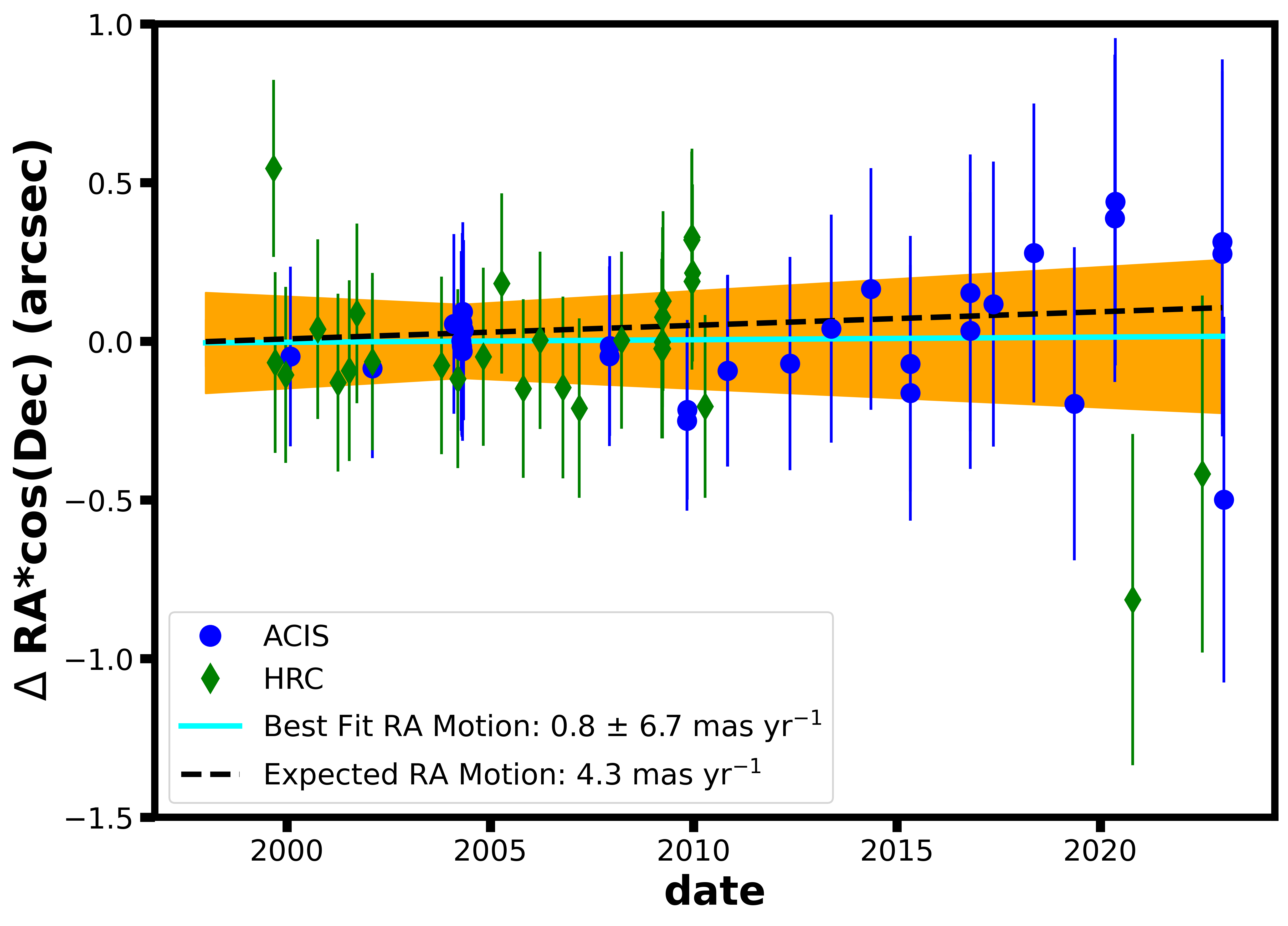}
\includegraphics[width=0.48\textwidth]{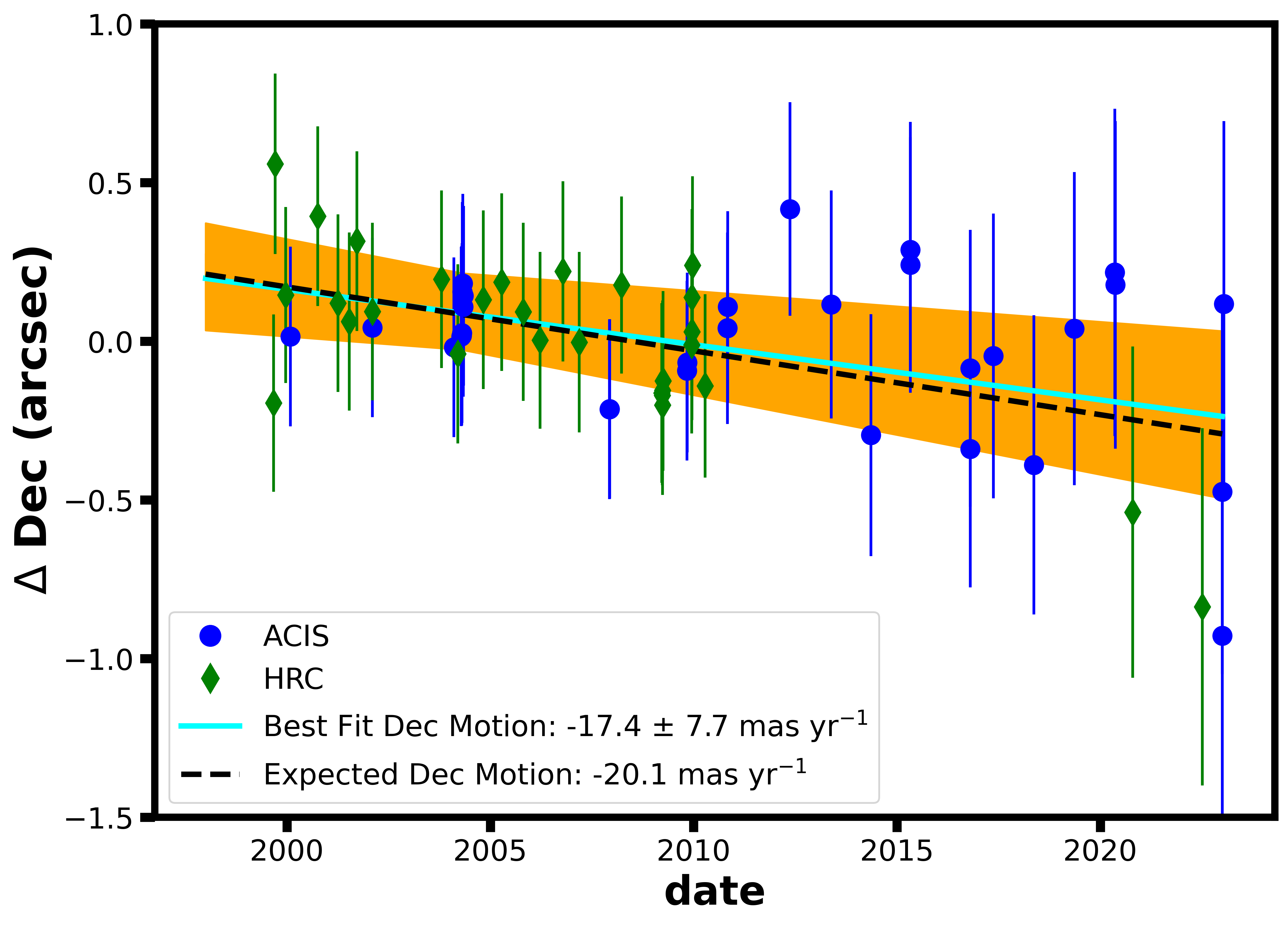}
\end{center}
\vspace{-5mm}
\caption{\footnotesize{The measured RA \& Dec offsets of the Cas~A CCO from the average position. The 32 blue circles are from {\sc ACIS} observations and the 29 green diamonds are from {\sc HRC} observations. The error bars on each data point reflect the 1-$\sigma$ total uncertainty (\texttt{wavdetect} $+$ astrometric). The black dashed line shows the expected NS motion found by back-evolving the NS to the explosion site identified by \cite{thorstensen01}, extrapolated from combined 2004 observations. The solid cyan line shows the best-fit motion, and the orange curve shows the 1-$\sigma$ likelihood region for the line of best-fit, including uncertainty in both slope and intercept and taking the average 2004 NS location as the reference point.}}
\label{fig:NS_motion}
\end{figure*}

After obtaining the PSF files, we followed the steps in the ```Accounting for PSF Effects in 2D Image Fitting''' \texttt{CIAO} Sherpa \citep{freeman01} thread\footnote{\url{https://cxc.cfa.harvard.edu/sherpa/threads/2dpsf/}} to fit each point source using a Gaussian and a constant background and the PSF as the convolution kernel. For each point source in each observation, we obtained a best-fit centroid position along with uncertainties. Using the reported Gaia point source positions and proper motions, we evolved each point source from its 2016 Gaia position to the location it would be at in each {\it Chandra} observation. As Gaia point source uncertainties are small ($\lesssim$0.001''), we treated these positions as the ``true'' reference locations for our calibration point sources.

\subsection{Registration Method}
At this point, we had robust estimates of calibration point source centroids along with precise reference locations from Gaia for each observation. The next step was to correct the astrometry of each {\it Chandra} observation by solving for a transformation matrix such that, when applied to our observations, the identified point sources have positions as close as possible to the reported Gaia positions. The generic transformation matrix is given by
\begin{equation}
    \begin{pmatrix}
    x'\\
    y'
    \end{pmatrix}
    = 
    \begin{pmatrix}
    {\rm r} \cos\theta & -{\rm r} \sin\theta\\
    {\rm r} \sin\theta & {\rm r} \cos\theta
    \end{pmatrix}
     \begin{pmatrix}
    x\\
    y
    \end{pmatrix}
    +
     \begin{pmatrix}
    t_1\\
    t_2
    \end{pmatrix}
\end{equation}
where (x,y) are input pixel coordinates of point sources,  (x',y') are the transformed coordinates, r represents a linear stretching of the image, $\theta$ is rotation, and (t$_1$,t$_2$) represents translations. We restricted our transformation matrix to only allow for translation shifts, as the plate scale is well-known and feasible rotation errors will produce small positional changes (0.1$^\circ$ roll angle error corresponds to only a 0.1'' shift for sources 1' away) relative to the $\sim$0.5'' astrometric uncertainty. Additionally, some of our images only had 1 or 2 calibration point sources, not enough to find a solution to a full transformation matrix that allowed rotation and/or stretching.

We used a least squares algorithm (Python function \texttt{scipy.optimize.least\_squares}) to solve for the transformation matrix, weighting each point source by its centroid uncertainty. Specifically, we used the `\texttt{soft\_l1}' loss function option (\citealt{triggs00}; as used in \citealt{long22}) that weights outliers with a linear rather than a quadratic penalty. After obtaining the best-fit translation shifts for each observation, we calculated corrected NS positions by shifting the NS positions (also found via 2D fitting the observation convolved with the PSF) by those values.

\subsection{Uncertainties on Final NS Positions}
There are three sources of error to combine in quadrature: 1.) the best-fit centroid uncertainty of the NS, 2.) the overall uncertainty of the point sources used to calculate the astrometric solution, and 3.) the accuracy of the astrometric solution. The Gaia point source uncertainties are negligible compared to the point source and astrometric correction uncertainties ($\lesssim$0.001'' compared to $\gtrsim$0.05'').

The first uncertainty is simply reported by the process of fitting the point source centroid, as described in Section~\ref{subsec:psf}. The second and third sources of error result from our decision to weight each point source by its centroid uncertainties when calculating the transformation matrix. As the final translation shift is effectively a weighted average, the uncertainty of the transformation shifts is calculated by adding the inverse centroid errors in quadrature:
\begin{equation}
    \sigma_{\rm PSs, tot} = \left(\frac{1}{\sum_i^N (1/\sigma_{i}^2)}\right)^{0.5}
\end{equation}
We calculate this error separately for both RA \& Dec positions.

Finally, the accuracy of the final transformed image---the registration error---is calculated as a weighted average of the differences between the corrected point source locations and their expected Gaia positions (d), using the calibration point sources' centroid uncertainties as the weights:
\begin{equation}
    \sigma_{\rm Res, tot} = \frac{\sum_i^N (d/\sigma_{i}^2)}{\sum_i^N (1/\sigma_{i}^2)}
\end{equation}

\begin{figure*}
\begin{center}
\includegraphics[width=0.48\textwidth]{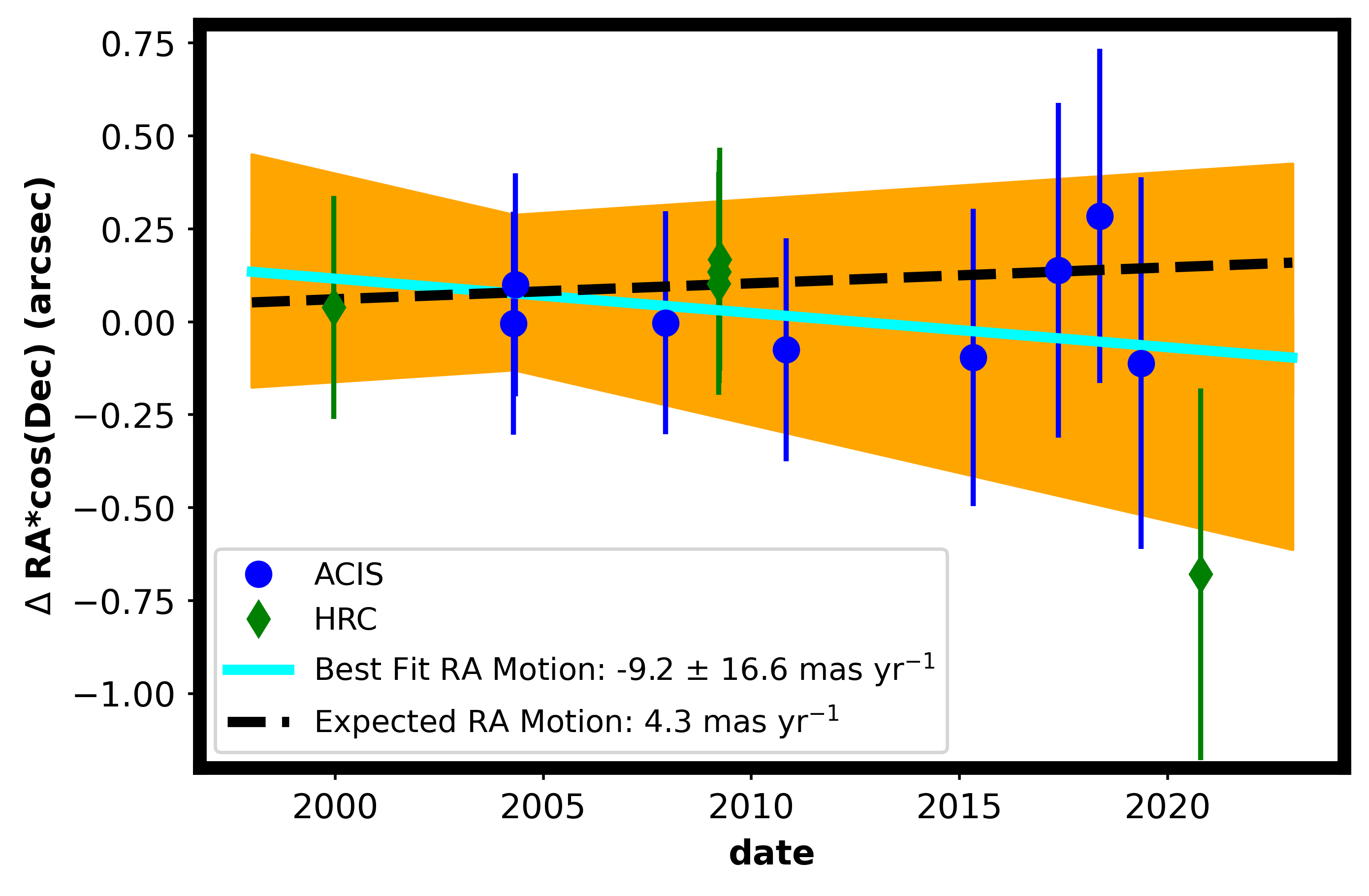}
\includegraphics[width=0.48\textwidth]{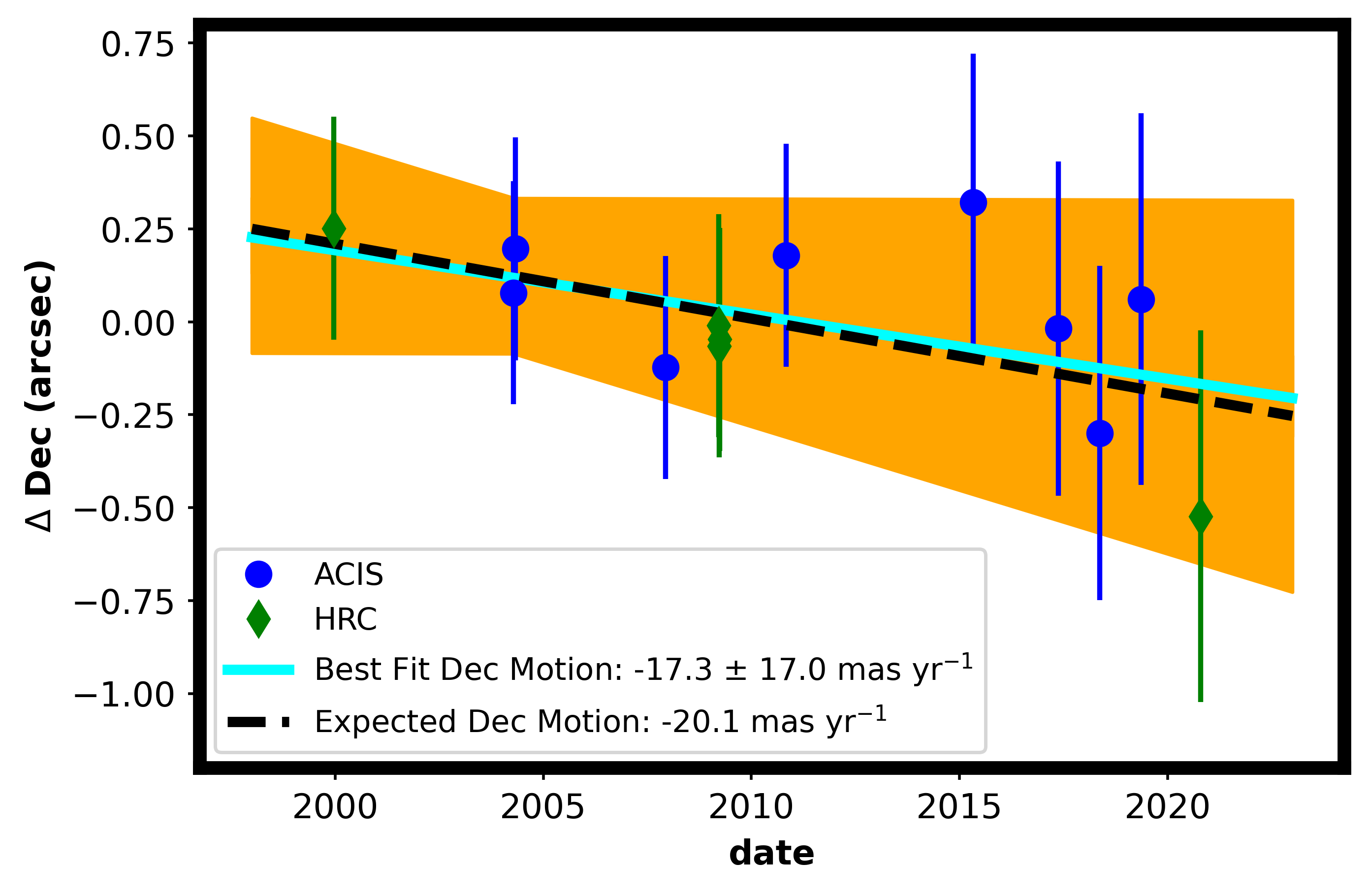} \\
\includegraphics[width=0.48\textwidth]{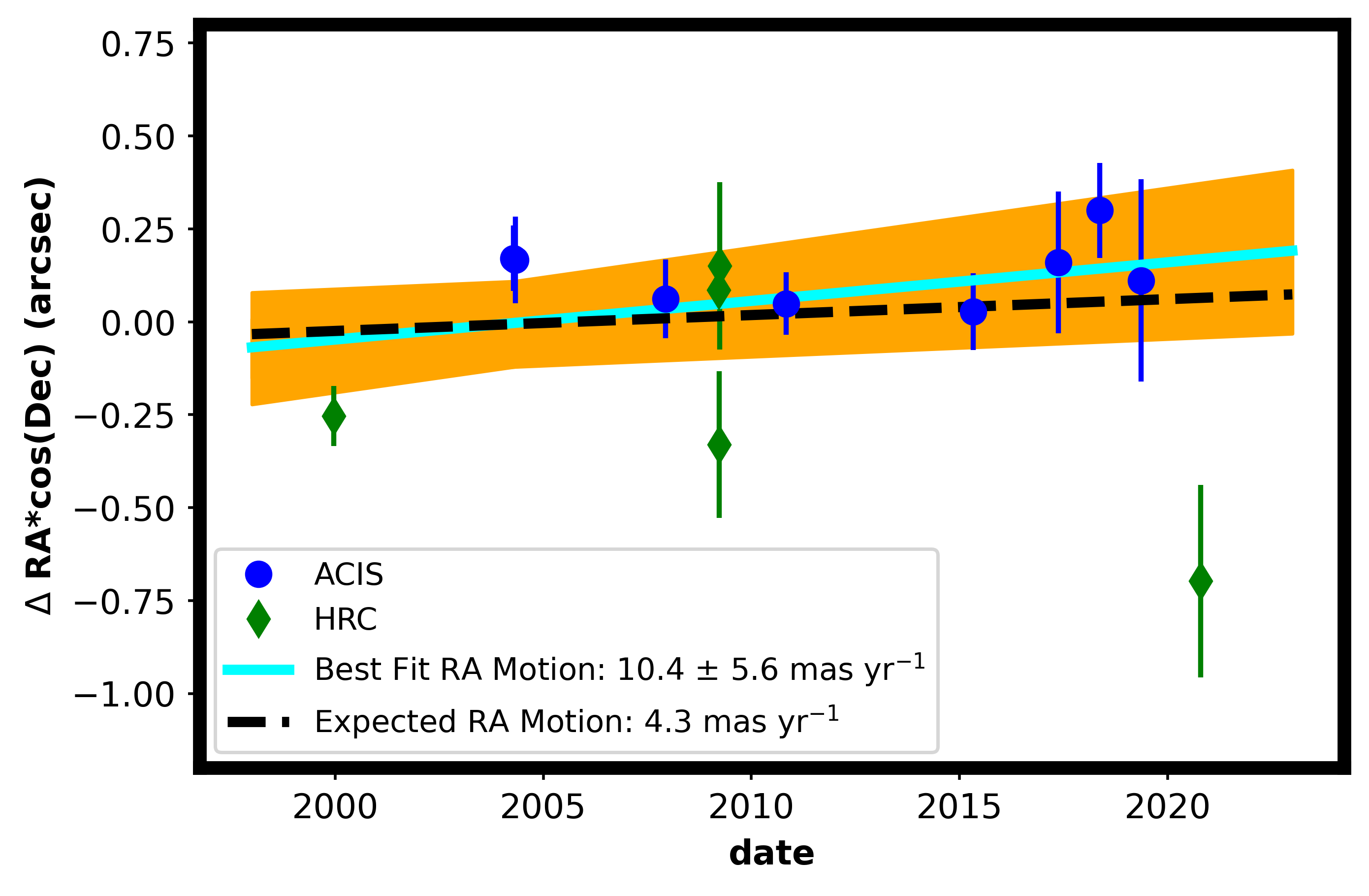} 
\includegraphics[width=0.48\textwidth]{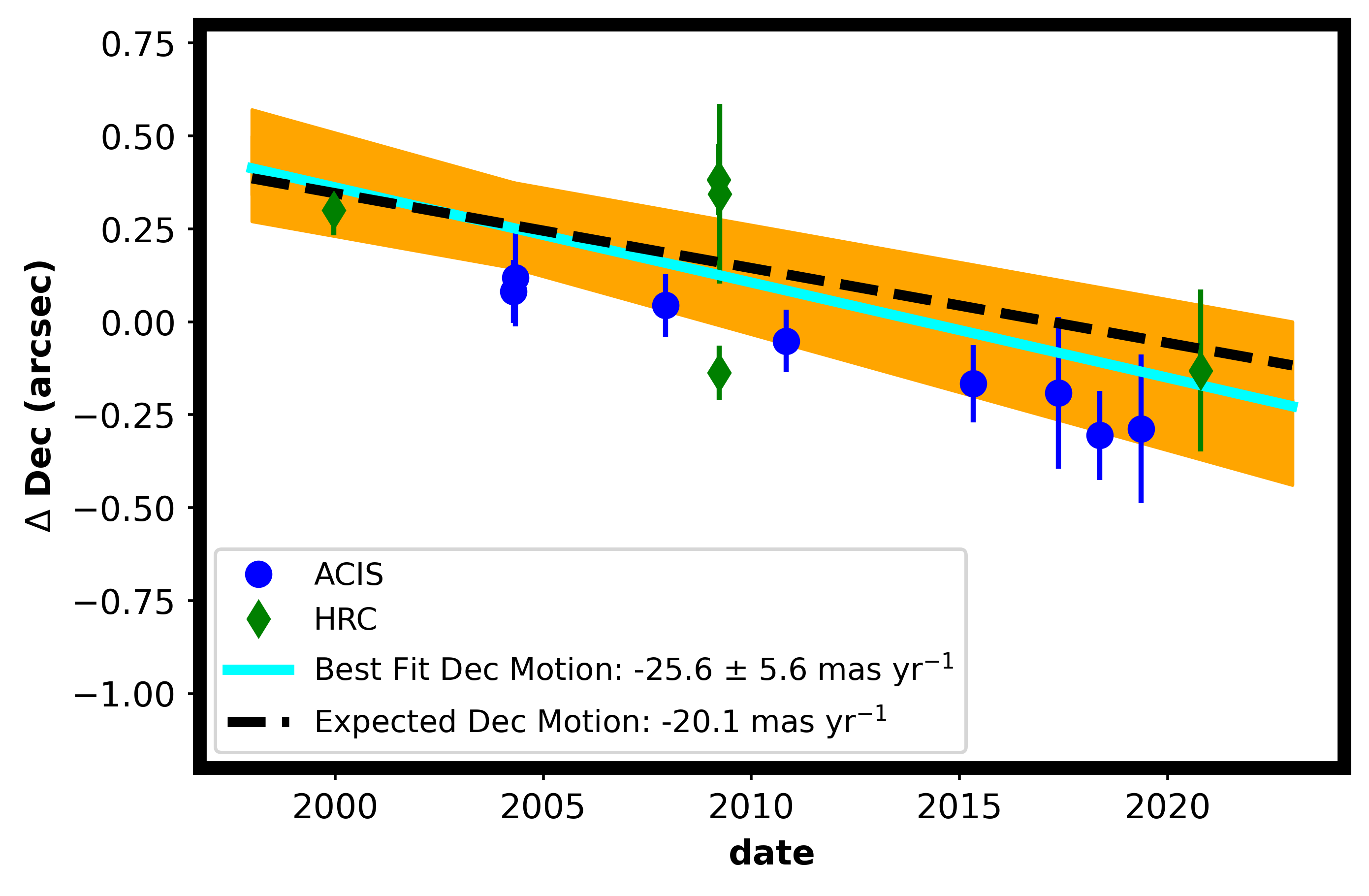} 
\end{center}
\vspace{-5mm}
\caption{\footnotesize{The measured RA \& Dec offsets of the Cas~A CCO from the average position using the 13 observations with identified calibration point sources: Top) no astrometric correction, Bottom) weighted astrometric correction. The black dashed line shows the expected NS motion based on the Cas A explosion site measurement by \cite{thorstensen01}, the solid cyan line shows our calculated best fit motion, and the orange curve shows the 1-$\sigma$ likelihood region for the line of best-fit, including uncertainty in both slope and intercept. All are calibrated using the 2004 NS location as the reference point. }}
\label{fig:NS_motion_corr}
\end{figure*}

\section{Proper Motion of the NS}
\label{sec:results}

\subsection{Simple Method: All Observations}
In the so called ``Simple Method'' we simply add the expected detector astrometric uncertainty (as estimated in Section~\ref{subsec:astrouncert}) to the reported best-fit centroid uncertainties of the NS at each epoch. We plot the NS locations measured from 32 {\sc ACIS} and 29 {\sc HRC} observations across 24 years and use \texttt{scipy.optimize.curve\_fit} \citep{vugrin07} in Python to perform non-linear least squares fitting to calculate the RA and Dec motion of the NS. .

We obtained the following proper motion measurements. RA: 0.8 $\pm$ 6.7 mas/yr; Dec: -17.4 $\pm$ 7.7 mas/yr. At a distance of 3.33~kpc, this corresponds to a final velocity of 275 km/s $\pm$ 121 km/s, with an angle of 177 $\pm$ 22 degrees east of north. The data and lines of best fit are shown in the top row of Figure~\ref{fig:NS_motion}.

\subsubsection{Simple Method Check}
To double check that our simple method is appropriate for use on {\it Chandra} data, we perform the same analysis on a well-known bright star that has many {\it Chandra} observations over the years and a robust Gaia proper motion measurement: AR Lac. We extracted the position of AR Lac in 34 Chandra HRC-I observations, applied the astrometric uncertainty to each of the data points, and calculated the line of best fit.

We obtained a measured motion for AR Lac of RA: -57 $\pm$ 8 mas/yr and Dec: 41 $\pm$ 8 mas/yr. The Gaia-reported proper motion is -52.31 $\pm$ 0.02 mas/yr in RAcos(Dec) and 46.83 $\pm$ 0.02 mas/yr in Dec. These results are within 1-$\sigma$ of each other, indicating that this technique of using dozens of {\it Chandra} images across multiple decades can enable the accurate measurement of point source proper motions.

\subsection{Astrometry-corrected Results}
\label{sub:as-corr}
Using our astrometric correction method on the 13 observations, we find a final RA motion of 10.4 $\pm$ 5.6 mas/yr and a final Dec motion of -25.6 $\pm$ 5.6 mas/yr. At a distance of 3.33~kpc, this corresponds to a total velocity of 436 $\pm$ 89 km s$^{-1}$ at an angle of 158 $\pm$ 12 degrees east of north. Figure~\ref{fig:NS_motion_corr} presents the NS positions from these 13 observations, showing proper motions calculated both without and with astrometric corrections.

\subsection{Local Rest Frame Velocity}
\label{sub:gal-corr}
One final step is to correct for Galactic rotation and the peculiar motion of the sun to find the motion of the neutron star in its' local rest frame. We use a solar distance to the Galactic center of 8.2~kpc, a solar peculiar velocity of v$_{ \odot}$ = (8.0, 12.4, and 7.7) km s$^{-1}$, and a Galactic rotation speed v$_{\rm LSR}$ of 236 km s$^{-1}$ with a flat rotation curve at these distances \citep{gawa19}. Based on the trigonometry framework presented in the appendices of \cite{verbunt17} we calculate these corrections and find an apparent -2.89 mas yr$^{-1}$ (-45.8 km s$^{-1}$) motion in RA and -1.66 mas yr$^{-1}$ (-26.2 km s$^{-1}$) motion in Dec. % for a total velocity of 52.7 km s$^{-1}$. 
We then subtract these values from our observed proper motions to get the true proper motion of the NS.

With our ``Simple Method,'' we find a corrected total velocity of 256 km s$^{-1}$ at an angle of 167$^\circ$ east of north. With our astrometry-correcting method, we find %motions of 13.3 mas yr$^{-1}$ (210 km s$^{-1}$) in RA and -23.9 mas yr$^{-1}$ (378 km s$^{-1}$) in Dec for 
a total velocity of 433 km s$^{-1}$ at an angle of 151$^\circ$.

\section{Discussions and Conclusions}
\label{sec:disc}

In this paper, we have taken advantage of the greater-than-two-decades' worth of {\it Chandra} observations to measure the proper motion of the neutron star in Cassiopeia~A. Using a straightforward method that doesn't correct for astrometric uncertainty but simply accounts for it as an added uncertainty on the NS position across 61 {\sc ACIS} and {\sc HRC} {\it Chandra} observations, we find an observed proper motion corresponding to a velocity of 275 km/s $\pm$ 121 km/s, with an angle 177 $\pm$ 22 degrees east of north for a distance of 3.33~kpc. Using a more complicated method where we correct 13 observations' astrometric solutions via associating detected point sources with known Gaia sources, we obtain a velocity of 436 $\pm$ 89 km s$^{-1}$ at an angle of 158 $\pm$ 12 degrees. These are observed heliocentric proper measurements; correcting for galactic rotation and the sun's peculiar motion decreases the estimated velocities by 5--20 km s$^{-1}$ and the estimated kick direction by $\sim$10$^\circ$.

Our results are separated by 1.08-$\sigma$ and both are within $\sim$1-$\sigma$ of the back-evolved explosion site velocity of $\sim$340 $\pm$ 30 km s$^{-1}$ \citep{thorstensen01} and the 570 $\pm$ 260 km s$^{-1}$ proper motion measurement using a baseline of 10--years \citep{mayer21}. The uncertainties of our estimates are smaller by a factor of 2--3 which we attribute to the longer baseline and use of more observations. 

Additionally, both of our results produce smaller velocity estimates than found previously. Our astrometry-uncorrected estimate excludes velocities $>$597 km s$^{-1}$ at 99\% confidence ($\gtrsim$482 km s$^{-1}$ at 90\% confidence), and our astrometry-corrected estimate excludes velocities $>$677 km s$^{-1}$ at 99\% confidence ($\gtrsim$593 km s$^{-1}$ at 90\% confidence). This phenomenon, where velocity estimates are revised to lower velocities when using longer baselines, has been found in proper motion studies of other SNR-embedded NSs. For example, with a 5-year {\it Chandra} baseline, the NS in Puppis~A was measured to have a velocity of 1,122 $\pm$ 360 km s$^{-1}$ \citep{hui06} or $\sim$1,600 km s$^{-1}$ \citep{winkler07}. With a $\sim$19~year baseline, this measurement was brought down to 763 $\pm$ 63 km s$^{-1}$ \citep{mayer20}. We attribute this to the small sky motions---close to zero arcseconds per year---of NSs compared to pixel sizes, telescope astrometry, and other sources of uncertainty. Thus, measurements with short baselines become dominated by any uncertainties or systematic offsets and can become biased away from zero. 
%We attribute this to the small sky motions of NSs---close to zero arcseconds per year---such that short baseline measurements will be dominated by small uncertainties or systematic offsets that can bias a measurement away from zero.
Determining the true fraction of SNR-embedded NSs with velocities of $\gtrsim$1000 km s$^{-1}$ is essential for informing simulations and properly understanding explosion processes in SNe, and this trend suggests that we should be cautious in treating extremely high NS velocities as accurate unless long baselines are used and astrometric uncertainties are properly accounted for. %: e.g., the velocity of 1200 $\pm$ 300 km/s reported by Ho et al. (2020) https://ui.adsabs.harvard.edu/abs/2020MNRAS.498.4396H/abstract.}

We can compare our measured velocities to the results of 3D CCSNe simulations performed by \cite{burrows23}. According to their Figure~1., a kick velocity of $\sim$370 $\pm$ 71 km s$^{-1}$ (the weighted average of our local rest frame velocity estimates) roughly corresponds to a progenitor mass range of $\sim$15--20M$_\odot$. This matches well with its classification as a Type IIb SN \citep{krause08} with expected progenitor masses of 13--20M$_\odot$.

Finally, our proper motion measurements support previous studies on the relation between the NS kick and properties of the Cas~A supernova remnant. The magnitude of our measured kick velocity matches with the 460 km s$^{-1}$ values reproduced in Cas~A simulations by \citep{wongwathanarat17} using their ``Gravitational-Tugboat Mechanism'' where the NS is accelerated due to gravitational forces from asymmetric ejecta. Our estimated proper motion angle of $\sim$165$^\circ$ east of north is nearly opposite the reported bulk ejecta motion angle of 10--15$^\circ$ \citep{me17,katsuda18} and is directly opposite to the heaviest ejecta (Ar, Ca, Ti, \& Fe) angle of 340--360$^\circ$ \citep{grefenstette14,wongwathanarat17,me20a,picquenot21}, for opening angles of 150--155$^\circ$ and 165--180$^\circ$ respectively and supporting the conservation of momentum with ejecta scenario for generating NS kicks.

Using proper motion measurements that are corrected for galactic rotation, we obtain similar velocities but smaller opening angles of $\sim$140-145$^\circ$ (bulk ejecta motion) and $\sim$155-175$^\circ$ (heavy elements' motion): still consistent with the velocities reproduced in Cas~A simulations and nearly directly opposite the heaviest ejecta.

Our uncertainty is dominated by the later-epoch {\sc ACIS} and {\sc HRC} observations. In particular, the 2020 {\sc HRC} observation \#24840 was only 14~ks which---in combination with the effective area degradation of {\it Chandra} at low energies---led to only a single detected calibration point source. In fact, three of the four observations taken after 2016 (ObsIDs \#19604, 19606, and 24840) only have a single detected calibration point source. As shown in Figures~\ref{fig:NS_motion} and \ref{fig:NS_motion_corr}, the RA of the NS in observation \#24840 observation is a notable outlier. We note that there is currently a 35~ks {\it Chandra} {\sc HRC} proposal for this NS waiting to be taken, proposed for by Dr. Heinke. If this observation occurs and multiple Gaia point sources are detected, a more precise proper motion measurement could be made.

Finally, future X-ray telescopes with $\sim$arcsecond precision would greatly help constrain the velocity of this, and other, NSs associated with SNRs. Observations in the 2030s with such an instrument would provide a 30-year baseline compared to early {\it Chandra} observations, and in particular smaller off-axis PSFs would enable easier and more precise detections of fore- and background point sources with which to correct image astrometry. 
In the existing observations of Cas~A, we were unable to detect nearby point sources $\gtrsim$3' off-axis in {\it ACIS} observations, and similarly-distant point sources in {\it HRC} observations had large centroid uncertainties of up to $\sim$0.5''.

It has been shown---through this study and others---that point source centroid uncertainties {\it smaller} than the pixel size or PSF of a detector can be obtained. Thus, even telescopes with up to a few arcsecond spatial resolution should be sufficient to measure NS positions with sufficiently long, multi-decade baselines. The potential probe mission {\it AXIS}, the future proposed missions {\it Athena} and/or {\it Lynx}, and other X-ray telescopes with less than a couple arcsecond resolution are vital to continue our investigations of NS velocities. Until then, continued {\it Chandra} observations of select targets (e.g., G11.2-0.3, G18.9-1.1, Cas~A, MSH 11-62, and RCW~103) that were observed in the 2000s can enable astronomers to expand the population of SNR-embedded NSs with robust proper motion measurements.

\acknowledgments

Dr. Patrick Slane acknowledges support from NASA Contract NAS8-03060. Dr. Xi Long acknowledges support from CXC grants SP8-19002X and GO9-20068X and GO2-23045X, and NASA grant 80NSSC18K0988. The authors thank Dr. Tom Aldcroft for helpful discussions on Chandra astrometry and Dr. Vinay Kashyap for providing AR Lac data.

This paper employs a list of Chandra datasets, obtained by the Chandra X-ray Observatory, contained in~\dataset[Chandra Data Collection (CDC)  188]{https://doi.org/10.25574/cdc.188}. This research has made use of data obtained from the {\it Chandra Data Archive}, software provided by the {\it Chandra X-ray Center} (CXC) in the application packages CIAO (4.13, \citealt{fruscione06}) and Sherpa (4.13.0, \citealt{sherpa,freeman01}), the Python library SciPy \citep{scipy20}, and the HEASARC software package Xspec (v12.12.0; \citealt{arnaud96}). This work made use of data from the European Space Agency (ESA) mission {\it Gaia} (\url{https://www.cosmos.esa.int/gaia}), processed by the {\it Gaia} Data Processing and Analysis Consortium (DPAC, \url{https://www.cosmos.esa.int/web/gaia/dpac/consortium}). Funding for the DPAC has been provided by national institutions, in particular the institutions participating in the {\it Gaia} Multilateral Agreement.

\bibliography{CasA_NS_Vel}

\end{document}